\documentclass[aps,preprintnumbers,twocolumn,showpacs]{revtex4}
\usepackage{amsfonts,amssymb,amsmath,mathbbol}              
\usepackage[]{graphics,graphicx,epsfig,ulem}            

\begin{document}

\title{Spatially selective loading of an optical lattice
by light--shift engineering using an auxiliary laser field}
\date{\today}
\author{P. F. Griffin, K. J. Weatherill, S. G. MacLeod, R. M. Potvliege, and C. S. Adams}
\affiliation{Department of Physics, University of Durham,
Rochester Building, South Road, Durham DH1 3LE, England.}

\begin{abstract}

We report on a method of light--shift engineering where an
auxiliary laser is used to tune the atomic transition frequency.
The technique is used to selectively load a specific region of an
optical lattice. The results are explained by calculating the
differential light--shift of each hyperfine state. We conclude
that the remarkable spatial selectivity of light--shift
engineering using an auxiliary laser provides a powerful technique
to prepare ultra-cold trapped atoms for experiments on quantum
gases and quantum information processing.

\end{abstract}

\pacs{39.25.+k, 31.10.+z, 32.60.+i, 03.67.-a, 03.75.-b, 42.62.-b}


\maketitle

Optical dipole traps and optical lattices are finding an ever
increasing range of applications in experiments on Bose-Einstein
condensation (BEC) \cite{barrett,Weber,Ybbec,weiss}, optical
clocks \cite{katori}, single--atom manipulation
\cite{schl02,mcke03}, and quantum information processing (QIP)
\cite{bloch,meschede}. In many applications, one is interested not
only in the light--shift of the ground state, which determines the
trap depth, but also the relative shift of a particular excited
state. One has some control over this differential shift as the
ground and excited states have different resonances, and the laser
can be tuned to a `magic' wavelength where the ground and excited
state polarizabilities are the same \cite{katori,mcke03,sterr}.
However, using a `magic' wavelength is not always appropriate
either because one is no longer free to select the laser
wavelength to minimize spontaneous emission, or because of the
unavailability of suitable light sources.

In this paper, we report on a different method of light--shift
engineering where a second laser field is used to control the
excited state polarizability. One such method for controlling
ground--state hyperfine polarizabilities is discussed in
Ref.~\cite{kaplan}. Light--shift engineering using an independent
laser could have significant advantages over the use of a `magic'
wavelength for some applications, as it can be applied to specific
regions of a trap. For example, we show how the technique can be
used to selectively load a well defined region of an optical
lattice. Specifically, we consider the case of loading $^{85}$Rb
atoms into a deep CO$_2$ laser lattice. A quasi-electrostatic
lattice based on a CO$_2$ laser (wavelength 10.6~$\mu$m) is
particulary attractive for trapping cold atoms or molecules as it
combines low light scattering \cite{thomas} with a lattice
constant sufficiently large to allow single-site addressability
\cite{sche00}.
We show that by focussing an additional laser (a Nd:YAG laser with
wavelength 1.064 $\mu$m) on a specific region of the lattice we
can selectively load only into sites in this region. This
spatially selective loading is most effective when the cooling
light is blue--detuned relative to the unperturbed atomic
resonance. We explain the effect by calculating the differential
light-shifts between the ground and excited states in the presence
of two light fields. We show that only in the light--shift
engineered region where the CO$_2$ and Nd:YAG laser beams overlap,
is the differential shift negative allowing efficient laser
cooling. In addition, we show that for red-detuned cooling light,
light--shift engineering produces a significant enhancement in the
number of atoms loaded into a deep optical trap.

The experimental set-up employed to demonstrate controlled loading
by light--shift engineering is shown in Fig.~1. An octagonal
vacuum chamber, fitted with home-made Zinc Selenide (ZnSe) UHV
viewports \cite{cox} to accommodate the CO$_2$ laser beams,
provides a background pressure of 1.2$\times 10^{-10}$ Torr. A
focussed Nd:YAG laser beam, with variable power, locally heats an
alkali metal dispenser to provide a controllable source of thermal
$^{85}$Rb atoms \cite{grif05}. A CO$_2$ laser beam, (propagating
along the $z$--axis in Fig.~1) with power 45~W, is focussed to
form a waist ($1/{\rm e}^2$ radius) of 70~$\mu$m at the center of
the chamber. The beam is collimated and retro--reflected to form a
1D optical lattice. The intensity of the CO$_2$ laser is
controlled using an acousto-optic modulator (AOM). The intensity
at the center of the lattice, $I_0=2.3\times10^{6}$~Wcm$^{-2}$,
gives a ground state light--shift $U_0 = -\textstyle{1\over
2}\alpha_0I_0/(\epsilon_0c) =h(-36~{\rm MHz})$, where
$\alpha_0=335~a_0^3$ is the ground--state polarizability at
10.6~$\mu$m in atomic units. A Nd:YAG laser beam (propagating at
$+45^\circ$ to the $y$ axis in the $xy$ plane) with power 7.8 W is
focussed by a $f=150$~mm lens to overlap with the CO$_{2}$ lattice
in the trapping region. The Nd:YAG laser has a circular focus with
a beam waist of $30~\mu$m in the overlap region. This gives an
intensity of $I_0=5.5\times10^5$~Wcm$^{-2}$ leading to a ground
state light--shift $U_0 =h(-18.6~{\rm MHz})$ using our calculated
value of $\alpha_0=722~a_0^3$ at 1.064~$\mu$m. The CO$_2$ and
Nd:YAG laser beams are linearly polarized along the $x$ and $z$
axes, respectively.

\begin{figure}[ht]
\begin{center}
\includegraphics[width=8.5cm]{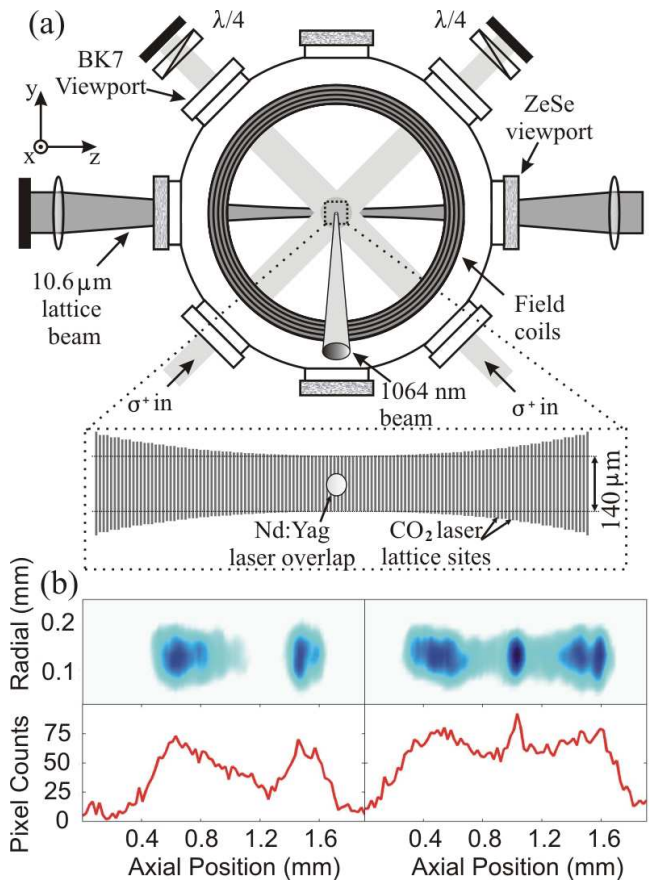}
\caption[]{(a) Experimental arrangement showing the intersection
of the CO$_{2}$ and Nd:YAG laser beams. Inset: Schematic
indicating the relative scales of the CO$_{2}$ and Nd:YAG spot
sizes in the interaction region. (b) Images and line profiles
without (left) and with (right) the Nd:YAG laser. The molasses
detuning (-50~MHz) is chosen to optimize the total number of atoms
rather than the number in the overlap region.
The viewing direction is approximately perpendicular to both the
CO$_2$ and Nd:YAG laser propagation directions.} \label{fig:1}
\end{center}
\end{figure}

Loading of a CO$_{2}$ laser lattice is carried out as follows: The
CO$_{2}$ and Nd:YAG laser beams are left on throughout the loading
stage. We load a magneto-optical trap (MOT), centered on the
dipole trap, with $2\times 10^{7}$ $^{85}$Rb atoms in typically 3
seconds. After the magnetic field is switched off, the cooling
laser beam intensities are reduced from 55~mWcm$^{-2}$ to
10~mWcm$^{-2}$ and the detuning is increased to $\Delta=-8\Gamma$,
where $\Gamma=2\pi(6~{\rm MHz})$ is the natural linewidth of the
transition, to create an optical molasses. After 10~ms of
molasses, the atom cloud has a typical temperature of 40~$\mu$K,
measured by time--of--flight. During the molasses phase the
hyperfine repumping laser intensity is lowered from 6~mWcm$^{-2}$
to 200~$\mu $Wcm$^{-2}$ and then switched off completely with a
shutter for the final 5~ms such that atoms are pumped in the lower
hyperfine state \cite{adam95}. After the molasses phase, the
cooling light and the Nd:YAG laser are extinguished for a few
hundred milliseconds, then the CO$_2$ laser is turned off and the
MOT beams (tuned to resonance) are turned back on to image the
cloud. A CCD camera collects the fluorescence to give a spatial
profile of the trapped atoms. A typical atom distribution viewed
approximately perpendicular to the CO$_2$ and Nd:YAG beam axes is
shown in Fig.~1(b). One sees that the CO$_2$ lattice loads
efficiently out in the wings where the trap depth is smaller. This
effect has been widely observed in experiments on far-off
resonance optical dipole traps \cite{kupp00,sche00,cenn03} and
arises due to the smaller differential light--shift between the
ground and excited states away from the focus. We also see that
the loading is greatly enhanced in the region where the Nd:YAG
laser intersects the CO$_2$ laser lattice.

Remarkably, if we detune the cooling light slightly to
the blue of the unperturbed atomic resonance such that
neither the CO$_2$ nor the Nd:YAG laser beams
alone trap any atoms,  then we still observe that the region where the
two beams intersect is efficiently loaded, see Fig. 2(b).
This spatial selectivity provides a very clear demonstration of the power
of light--shift engineering using an auxiliary laser field.
In addition, it demonstrates that the enhanced loaded observed in Fig.~1 and Fig.~2(a)
cannot be explained by a `dimple' effect \cite{dimple},
where atoms from the wings of a trap rethermalise in
the deeper overlap region \cite{taku03}.

\begin{figure}[ht]
\begin{center}

\vspace*{0.5cm}

\includegraphics[width=8.5cm]{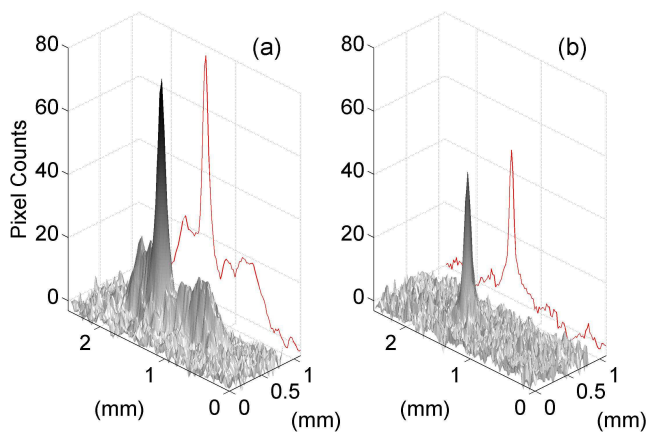}
\caption[]{A surface plot of the column density for cooling laser
detunings (a) $\Delta=2\pi(-20$~MHz) and (b)
$\Delta=2\pi(+2$~MHz). The on--axis density is shown on the back
plane. For blue--detuning (b) only the light--shift engineered
region, where the CO$_2$ and Nd:YAG laser beams overlap, is
loaded.} \label{fig:2}
\end{center}
\end{figure}

Finally, we should add that the enhanced loading observed in the overlap region cannot
be explained simply by the fact that the trap is
deeper in this region. To demonstrate this
we have reduced the CO$_2$ laser power by a factor of four
such that the depth in the combined CO$_2$ plus Nd:YAG trap is less than
a CO$_2$ lattice alone at full power. Typical column densities
are shown in Fig. 3. We see that the density in the
combined trap is still significantly higher than for a deeper CO$_2$ lattice.

\begin{figure}[ht]
\begin{center}
\includegraphics[width=7.5cm]{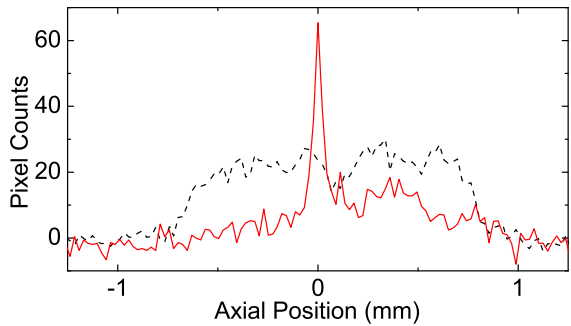}
\caption[]{Column density for a CO$_2$ laser lattice without the
Nd:YAG laser (dashed line), and for a shallower CO$_2$ laser
lattice with the Nd:YAG laser (solid line). The overall ground
state light--shift in the overlap region of the shallow combined
trap ($-27$~MHz) is less than the maximum light-shift for the
CO$_2$ laser lattice alone ($-36$~MHz), but loading into the
combined trap is still significantly more efficient. Both profiles
are for a molasses detuning of $-20$~MHz.} \label{fig:3}
\end{center}
\end{figure}

To explain the spatially selective loading for blue--detuned
cooling light, we have calculated the polarizability of the $5s$
ground and $5p$ excited states as a function of wavelength. The
details of the calculation will be explained elsewhere
\cite{potv05}. Briefly, the scalar polarizability $\alpha_0$ is
the average of the dipole polarizabilities $\alpha_{xx}$,
$\alpha_{yy}$ and $\alpha_{zz}$ for an atom exposed to a laser
field polarized, respectively, in the $x$, $y$, and
$z$-directions:
$\alpha_0=(\alpha_{xx}+\alpha_{yy}+\alpha_{zz})/3$. The scalar
polarizability is the same for all $m$-components of the $5p$
state. In addition, there is a tensor polarizabilty $\alpha
_{2}=\left(\alpha _{xx}-\alpha _{zz}\right)/{3}$ which lifts the
degeneracy of different $m$-states. In order to obtain these
quantities, we represent the interaction of the valence electron
with the core by the model potential proposed by Klapisch
\cite{Klapisch67}. The polarizabilities are calculated by the
implicit summation method \cite{implicit}. Thus $\alpha_{xx}$ (and
similarly for $\alpha_{yy}$ and $\alpha_{zz}$) is obtained as
$\alpha_{xx} = - e(\langle 0 \vert x \vert 1 \rangle +\langle 0
\vert x \vert -1 \rangle)/\mathcal{F}$, where $\vert 0 \rangle$
represents the state vector of the unperturbed $5s$, or
$5p_{-1,0,1}$ states, and $\vert\pm 1 \rangle$ are such that
\begin{equation}
(E_0 \pm \hbar\omega - H_0)\vert\pm 1 \rangle = e\mathcal{F}x \vert 0 \rangle.
\end{equation}
Here $H_0$ is the Hamiltonian of the field-free model atom and
$E_0$ is the eigenenergy of the unperturbed state, i.e.
$H_0 \vert0 \rangle = E_0 \vert0 \rangle$, and $\mathcal{F}$ is
an arbitrary electric field.
These equations are solved in position
space by expanding the wave functions on a discrete basis of radial Sturmian
functions and spherical harmonics
\cite{Potvliege98}.
In the zero-frequency limit, the resulting values of $\alpha_0[5s]$,
$\alpha_0[5p]$ and $\alpha_2[5p]$ converge towards
$333 a_0^3$, $854 a_0^3$, and $-151a_0^3$, respectively, in satisfactory
agreement with previous experimental and theoretical work
\cite{Safronova99}.
The dynamic polarizabilities as functions of wavelength are shown
in Fig.\ 4. We find that $\alpha_0=722a_0^3$ at 1.064 $\mu$m,
which agrees well with experiment and other theoretical work
\cite{Bonin93,Marinescu94,Safronova04}.

\begin{figure}[ht]
\begin{center}
\includegraphics[width=7.5cm]{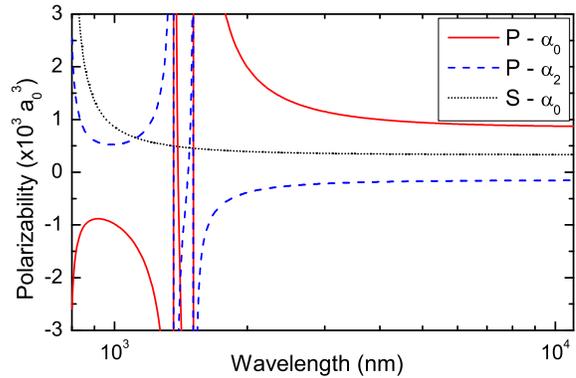}
\caption[]{Calculated polarizabilities of the $5s$ and $5p$ states
of Rb. For the $p$ state we show the scalar and tensor
polarizabilities, $\alpha _{0}=\left( \alpha _{xx} + \alpha
_{yy}+\alpha _{zz}\right)/{3}$ and $\alpha _{2}=\left(\alpha
_{xx}-\alpha _{zz}\right)/{3}$, respectively. For the $s$ state,
$\alpha _0= \alpha _{xx}= \alpha _{yy}= \alpha _{zz}$.}
\label{fig:4}
\end{center}
\end{figure}

\begin{figure}[ht]
\begin{center}
\includegraphics[width=8cm]{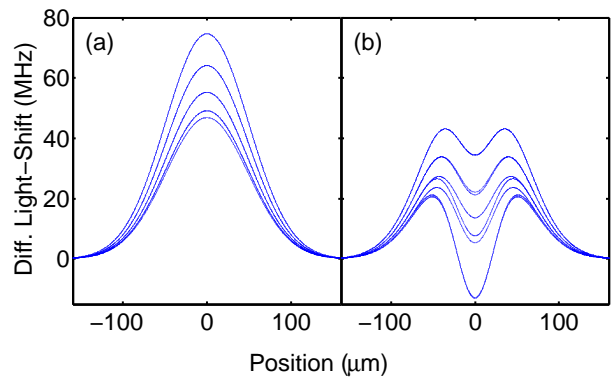}
\caption[]{The differential light--shifts as a function of
position along an axis perpendicular to both the CO$_2$ and Nd:YAG
laser propagation directions. The differential light--shift
corresponds to the additional detuning of the cooling laser seen
by ground--state atoms. It is equal to the
light--shifts of the $m_F=-4,\ldots,+4$ magnetic sub-levels of the
$5p^2P_{3/2}(F=4)$ minus that of the ground state state in
$^{85}$Rb for atoms in (a) the CO$_2$ laser lattice only, and (b)
in the combined CO$_2$ plus Nd:YAG trap. } \label{fig:5}
\end{center}
\end{figure}

For our purposes the most important result of Fig.~4 is that the
polarizabilities of the $5s$ state at the CO$_2$ laser wavelength
($\lambda=10.6~\mu$m) and the Nd:YAG wavelength
($\lambda=1.064~\mu$m) have the same sign, whereas the
polarizabilities of the $5p$ state have opposite signs. It follows
that one can use a combination of CO$_2$ and Nd:YAG lasers to tune
the differential light--shift between the $5s$ and $5p$ states
through zero. To calculate the light--shift experienced by atoms
in the combined CO$_2$ plus Nd:YAG trap we calculate the
eigenvalues of the matrix \cite{ange68,schm73}
\begin{eqnarray}
\mathsf{U} & = & \mathsf{U}_0 -\frac{1}{2\epsilon_0c}
\sum_{i=1,2}
(\alpha_0^i \mathbb{1} + \alpha_2^i \mathsf{Q}^i)I_i~,~
\end{eqnarray}
where $\mathsf{U}_0$ is a diagonal matrix with components
corresponding to the hyperfine splitting, the index $i$ denotes
the CO$_2$ and Nd:YAG lasers, and $\mathsf{Q}^i$ is a matrix with
components $\langle F,m_F\vert Q_\mu\vert F',m_F'\rangle$ with
$Q_\mu=[3\hat{J}^2_\mu-J(J+1)]/J(2J-1)$ and $J_\mu$ being the
electronic angular momentum operator in the direction of laser
field $i$. The differential light--shift between the ground state
and the $5p^2P_{3/2}(F=4)$ state for the CO$_2$ laser alone is
shown in Fig.~5(a). As a single laser beam splits the states
according to the magnitude of $m_F$, there are five curves
corresponding to $\vert m_F\vert=0,\ldots,4$. We see that all the
levels are far blue-detuned (positive differential shift) at the
centre of the lattice, making laser cooling ineffective unless the
cooling light is detuning to the red by an amount larger than the
differential light--shift. Adding the Nd:YAG laser produces the
shifts shown in Fig.~5(b). The Nd:YAG laser lifts the degeneracy
between the $\pm m_F$ components, although two pairs of states
remain close to degenerate. More importantly, one pair of states
is pulled down into the region of negative differential shift.
This allows efficient laser cooling in the center of the overlap
region, even when the cooling light is slightly blue--detuned
relative to the unperturbed resonance frequency. Note that,
efficient loading for blue--detuning can only be explained if one
includes the tensor polarizability term $\alpha_2$. Although
$\alpha_2$ is smaller than the scalar polarizability (by a factor
of 4 or 5), it dramatically alters whether states see the cooling
light as red or blue detuned and therefore completely determines
whether the trap is loaded or not.

As light--shift engineering allows laser cooling to
work as efficiently as in free space one might expect to load atoms
at lower temperature than in conventional loading schemes.
To address this issue we need to increase the sensitivity and the resolution
of our imaging system to allow
accurate density and temperature measurements.
This will be the focus of future work.

To conclude, we have shown how light-shift engineering using
an auxiliary laser field can be used to implement spatially
selective loading of deep far-off resonance optical lattices.
We have performed theoretical calculations of the
atomic polarizabilities and have shown that the addition of a second laser
field induces a splitting of the excited state which is crucial
in determining the efficiency of loading into the combined trap.
The technique could be applied to load a single-site in 3D
CO$_2$ lattice, with the interesting
prospect of BEC in the limit of high trap frequency.
In addition one could adapt the technique to
perform patterned loading of optical lattices \cite{peil03}
for applications in QIP experiments.


We thank E. Riis, I. G. Hughes, and S. L. Cornish for stimulating discussions,
M. J. Pritchard for experimental assistance and EPSRC for financial
support.


\end{document}